# ARPES study of the epitaxially grown topological crystalline insulator SnTe(111)

Yi Zhang[1,2,3*], Zhongkai Liu[4,5], Bo Zhou[3,6], Yeongkwan Kim[3], Lexian Yang[3,6,7], Hyejin Ryu[3,8], Choongyu Hwang[8,9], Yulin Chen[4,6,7], Zahid Hussain[3], Zhi-Xun Shen[2,5], Sung-Kwan Mo[3*]

[1]*National Laboratory of Solid State Microstructures, School of Physics, Collaborative Innovation Center of Advanced Microstructures, Nanjing University, Nanjing 210093, China*

[2]*Stanford Institute of Materials and Energy Sciences, SLAC National Accelerator Laboratory, Menlo Park, CA 94025, USA*

[3]*Advanced Light Source, Lawrence Berkeley National Laboratory, Berkeley, CA 94720, USA*

[4]*School of Physical Science and Technology, ShanghaiTech University, Shanghai 200031, China.*

[5]*Geballe Laboratory for Advanced Materials, Departments of Physics and Applied Physics, Stanford University, Stanford, CA 94305, USA*

[6]*Department of Physics and Clarendon Laboratory, University of Oxford, Parks Road, Oxford, OX1 3PU, UK*

[7]*State Key Laboratory of Low Dimensional Quantum Physics, Collaborative Innovation Center of Quantum Matter and Department of Physics, Tsinghua University, Beijing 100084, China.*

[8]*Max Plank POSTECH Center for Complex Phase Materials, Pohang University of Science and Technology, Pohang 790-784, Korea*

[9]*Department of Physics, Pusan National University, Busan 609-735, Korea*

*Email：zhangyi@nju.edu.cn
            skmo@lbl.gov

**SnTe is a prototypical topological crystalline insulator, in which the gapless surface state is protected by a crystal symmetry. The hallmark of the topological properties in SnTe is the Dirac cones projected to the surfaces with mirror symmetry, stemming from the band inversion near the L points of its bulk Brillouin zone, which can be measured by angle-resolved photoemission. We have obtained the (111) surface of SnTe film by molecular beam epitaxy on $BaF_2(111)$ substrate. Photon-energy-dependence of *in situ* angle-resolved photoemission, covering multiple Brillouin zones in the direction perpendicular to the (111) surface, demonstrate the projected Dirac cones at the $\bar{\Gamma}$ and $\bar{M}$ points of the surface Brillouin zone. In addition, we observe a Dirac-cone-like band structure at the $\Gamma$ point of the bulk Brillouin zone, whose Dirac energy is largely different from those at the $\bar{\Gamma}$ and $\bar{M}$ points.**

## 1. Introduction

The topological classification of materials has become a central part of condensed matter physics research [1, 2]. This classification includes topological insulators (TIs) [3], topological superconductors [4], topological crystalline insulators (TCIs) [5], topological Dirac semimetals, and Weyl semimetals [6]. The key features of the topological materials are the robust Dirac fermions and gapless Dirac cones in their surface or bulk states, which are protected by the symmetries defining their topological nature. The manipulation of these Dirac fermions could lead to many novel phenomena in the topological quantum systems, such as the quantum spin Hall effect in HgTe quantum wells [7, 8], quantum anomalous Hall effect in magnetically doped TI systems [9], Majorana fermions in topological superconductors [10, 11]. These properties make the topological quantum materials good candidates for wide variety of applications in non-dissipative system, quantum computing and spintronic devices [12].

As a member of the topological quantum materials, TCIs host topological surface states protected by its point group symmetry of the crystal lattice rather than the time-reversal symmetry as in TIs [5, 13]. As a prototypical TCI, SnTe has a simple rock salt structure shown in Figure 1a, and its topological band inversions occur at the L points in the bulk Brillouin zone (BZ) (Figure 1b), which will induce the topological surface state in its high symmetry faces, such as (100) or (111) [13, 14]. Previous angle-resolved photoemission spectroscopic (ARPES) studies on SnTe(100) demonstrate a pair of Dirac cones around the $\bar{X}$ points of the surface BZ [15]. For SnTe (111) surface, two types of Dirac-cones, which are both originated from the band inversions at the L points in bulk BZ, are projected at the $\bar{\Gamma}$ point and $\bar{M}$ points of the (111) surface BZ [16-18]. A transition from topologically trival to non-trival insulators was also observed in both (100) and (111) surfaces upon Pb doping [19, 20], as well as in related material SnSe [21, 22]. There also have been efforts to introduce superconductivity in SnTe [23, 24], which is a key step for realizing the Majorana fermions in a condensed matter system [25].

Previous ARPES studies on SnTe(111), however, only focused on the topological surface state within limited regions in $k$-space, particularly in $k_z$ direction, i.e., only around photon energies that give the most pronounced signal for the topological surface state. A full photon-energy-dependent mapping of electronic structure across the bulk BZ of SnTe(111) is still lacking [16]. In this paper, we performed an ARPES study on the epitaxial SnTe(111) films with photon energies ranging from 21 eV to 110 eV, which allows us to observe the band structure from the Γ point to the L point in multiple bulk BZs in the direction perpendicular to (111) surface. In addition to the Dirac cones at the $\bar{\Gamma}$ point and $\bar{M}$ points of the surface BZ, consistent with the previously reported ARPES results, we observed yet another Dirac-cone-like band structure at the Γ point of bulk BZ, with a much lower Dirac energy position compared to the Dirac cones at the $\bar{\Gamma}$ and $\bar{M}$ points. The observation of such Dirac cone would call for more studies from both theory and experiment regarding the topological classification of SnTe.

2. Experimental

The experiments were performed at the Beamline 10.0.1, Advanced Light Source, Lawrence Berkeley National Laboratory. SnTe(111) thin film was obtained by molecular beam epitaxial (MBE) growth at the Beamline with a base pressure of ~2×10$^{-10}$ Torr. The samples were transferred to the analysis chamber (base pressure of ~3×10$^{-11}$ Torr) for the *in situ* ARPES measurements immediately after the MBE growth under ultra-high vacuum environment. In order to obtain the (111) surface of SnTe, a commercial BaF$_2$ wafer was first degassed at 400 $^o$C for 30 minutes as a substrate [26, 27]. The reflection high-energy electron diffraction (RHEED) patterns in Figures 1c & d show the sharp 1×1 diffraction spots from BaF$_2$(111) surface, indicating a high quality of the substrate [28, 29].

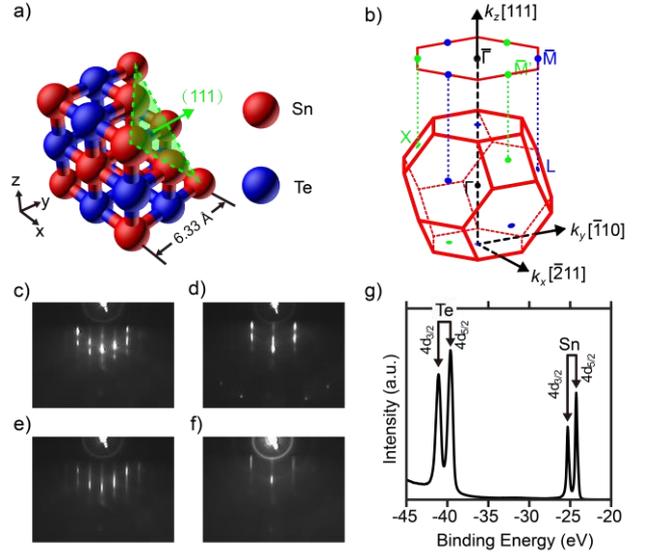

**Figure 1. MBE growth of SnTe(111). a)** Crystal structure of SnTe, the green plane and arrow indicate the (111) plane of SnTe. **b)** 3D bulk BZ of SnTe(111), and its projection on 2D surface BZ (upper hexagon). $k_x$, $k_y$, and $k_z$ denote the unit vectors in momentum space for (111) plane, along the direction of $\bar{\Gamma}$-$\bar{M}$ [$\bar{2}11$], $\bar{\Gamma}$-$\bar{K}$ [$\bar{1}10$], and [111], respectively. **c) & d)** RHEED patterns of BaF$_2$(111) substrate along the $\bar{\Gamma}$-$\bar{K}$ [$\bar{1}10$] and $\bar{\Gamma}$-$\bar{M}$ [$\bar{2}11$] directions, respectively. **e) & f)** RHEED patterns of grown SnTe(111) films along the $\bar{\Gamma}$-$\bar{K}$ [$\bar{1}10$] and $\bar{\Gamma}$-$\bar{M}$ [$\bar{2}11$] directions, respectively. **g)** Core level spectrum of SnTe(111) films.

The growth of the SnTe(111) film was achieved by evaporating high-purity (99.999%) SnTe crystals from a standard Knudsen cell at 480 °C, and depositing it onto the BaF$_2$(111) substrate. During the growth, the substrate was kept at ~250 °C for crystallization of SnTe(111) film. The deposition rate is about one unit-cell (UC) per one minute, monitored by the *in situ* RHEED system. Figures 1e & f show the RHEED patterns of the SnTe(111) film with thickness ~60 UC (~25 nm) along the $\bar{\Gamma}$-$\bar{K}$ [$\bar{1}$10] and $\bar{\Gamma}$-$\bar{M}$ [$\bar{2}$11] direction, respectively. The long 1×1 diffraction stripes indicate the high quality of [111] orientation of the SnTe film [20].

ARPES data were taken with a Scienta R4000 electron analyzer at 60K. The photon energies from 21 eV to 110 eV were used for the measurements, with energy and angular resolutions of 25 meV and 0.1°, respectively. The size of the beam spot on the sample was ~100 μm × 150 μm. Figure 1g is the angle-integrated core level spectrum taken by the *in situ* ARPES system. The sharp characteristic peaks of Sn (4d$_{5/2}$ orbit at 24.3 eV and 4d$_{3/2}$ orbit at 25.3 eV) and Te (4d$_{5/2}$ orbit at 39.6 eV and 4d$_{3/2}$ orbit at 41.1 eV) indicate the stoichiometry and purity of the SnTe film.

## 3. Results

Figures 2a-r are the photon-energy-dependent ARPES spectra of epitaxially grown SnTe(111) film with selected photon energies from 21 eV to 110 eV, along the $\bar{M}$'-$\bar{\Gamma}$-$\bar{M}$ direction of surface BZ. Figure 2s is the experimental Fermi surface (FS) obtained from ARPES with photon energies of 65 eV. The symmetry of the FS evidences the [111] orientation of the SnTe film, consistent

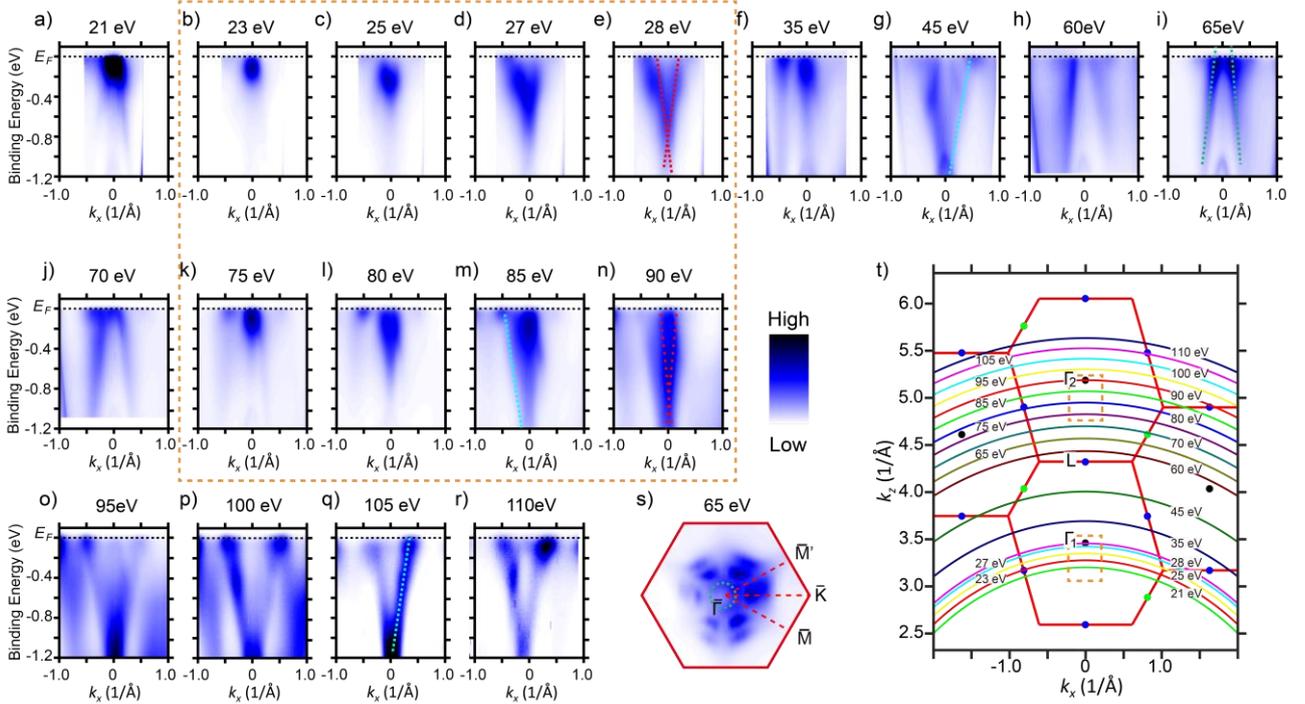

**Figure 2. Photon-energy-dependent ARPES spectra**. **a)-r)** Photon-energy dependent ARPES spectra of SnTe(111) film with selected photon energy from 21 eV to 110 eV, along the $\bar{M}$'-$\bar{\Gamma}$-$\bar{M}$ direction of surface BZ. The black dotted lines indicate the Fermi level. The red dotted lines in e) and n) depict the Dirac-cone-like linear band structure near the bulk Γ point. The green dotted lines in i) depict the linear bands of Dirac cone near the $\bar{\Gamma}$ point. The cyan dotted lines in g), m) and q) depict the linear bands of Dirac cones near the $\bar{M}$/$\bar{M}$' points. **s)** Fermi surface map of SnTe(111) film with photon energy of 65 eV. The red hexagon is the surface BZ of SnTe(111). The green dotted circle depicts the circular pocket near the $\bar{\Gamma}$ point. **t)** Photon-energy-dependent ARPES scanning traces in bulk BZ. The bronze dashed rectangles highlight the periodic replication of k)-n) to b)-e), and indicate the equivalent areas along the k$_z$ of bulk BZ, respectively.

with the RHEED patterns shown in Figure 1. We notice that the features of spectra with photon energies from 75 eV to 90 eV (Figures 2k-n) are essentially periodic replications of those with photon energies from 23 eV to 28 eV (Figures 2b-e), implying that the photon-energy-dependent ARPES spectra scan equivalent paths in the $k_z$ direction of the bulk BZ (bronze dashed rectangles in Figure 2). From these information, we can determine the positions of photon-energy-dependent ARPES scanning traces in the bulk BZ in Figure 2t with inner potential of $V_0$=21.5 eV [1]. For the spectra with 65 eV photon energy, a pair of linear band (green dotted lines in Figure 2i) becomes most visible, and a circular pocket is also observed in the Fermi surface (green dotted circle in Figure 2s). According to previous reports [16, 17], this linear band is the Dirac cone at the $\bar{\Gamma}$ point of (111) surface BZ projected from the band inversion at the L point of the bulk BZ. The linear bands of Dirac cones at the $\bar{M}$/ $\bar{M}$' points appear at the spectra with photon energy of 45 eV and 85 eV, and become most visible in the spectra of 105 eV photon energy (cyan dotted lines in Figures 2g, m & q). In our measurements, these linear bands of Dirac cones at the $\bar{\Gamma}$ and $\bar{M}$/ $\bar{M}$' points are found to be most pronounced, in terms of intensity, when the $k_z$ momentum in ARPES becomes close to the L point in the bulk BZ (Figure 2t). In Figure 3, we plot sliced constant energy maps along the $k_x$-$k_z$ plane. The bands of Dirac cones at the $\bar{\Gamma}$ point and $\bar{M}$/ $\bar{M}$' points (depicted by green and cyan dotted/dashed lines in Figure 3, respectively) show no dispersion along the $k_z$ direction.

In addition to the linear bands in the spectra of 65 eV, which we assign as topological surface states as in previous reports [16] and whose Dirac energy lies above the Fermi energy, we notice that there exist yet other Dirac-cone-like band structures with Dirac points below the Fermi energy in the spectra measured with 28 eV and 90 eV (red dotted lines in Figures 2e & n). These photon energies coincide with the $\Gamma_1$ and $\Gamma_2$ point of bulk BZ in Figure 2t, respectively. To enhance the visibility of these Dirac-cone-like features at the bulk Γ-points, we show the second-derivative spectra in Figures 4b & e, along with the detailed dispersion of these bands extracted by fitting momentum distribution curves (MDCs) in Figures 4c & f (green and red circles). The MDC peak positions exhibit highly linear behavior as shown in Figures 4a, b, d & e by green and red dotted lines. We determine the Fermi velocity and Dirac point of this emerging Dirac-cone-like band structure to be $v_F$ = 6.1±0.8×10$^5$ m/s and $E_D$ = -0.85±0.06 eV, respectively.

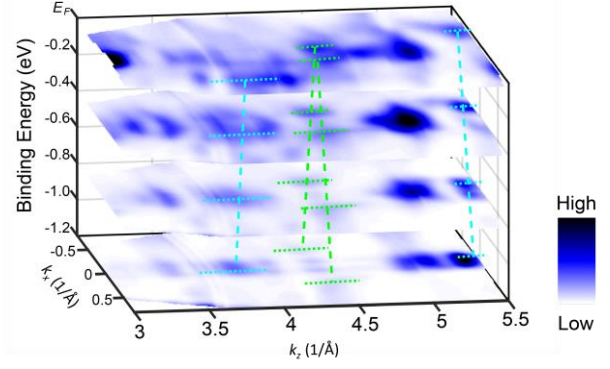

**Figure 3. Constant energy maps along the $k_x$-$k_z$ plane.** The green and cyan doted/dashed lines indicate the bands that make Dirac cones at the $\bar{\Gamma}$ and $\bar{M}$/ $\bar{M}$' points of (111) surface BZ, respectively.

Using the same method, the dispersion of the Dirac cones at the $\bar{\Gamma}$ and $\bar{M}$ points of the surface BZ can also be extracted from the MDCs. Figure 4h is the MDCs of the spectra with 65 eV photon energy. The green and red triangles indicate the peaks in MDCs, and the green and red dotted lines in Figure 4g are the linear fitting results, giving a Fermi velocity $v_F$ = 7.6±0.6×10$^5$ m/s and Dirac point position $E_D$=0.64±0.04 eV. The Dirac cone at $\bar{M}$ point becomes most visible in the spectra measured with 105 eV (Figure 4i), for which the scanning trace cutting through L point on the side in the bulk BZ. The MDC fitting gives the Fermi velocity $v_F$ ~ 4.9±1.2 × 10$^5$ m/s. Since the Dirac point at the $\bar{M}$ point is far above the Fermi level [16, 30], we can only roughly estimate the Dirac point position of $E_D$ = 1~1.2 eV.

4. **Discussion**

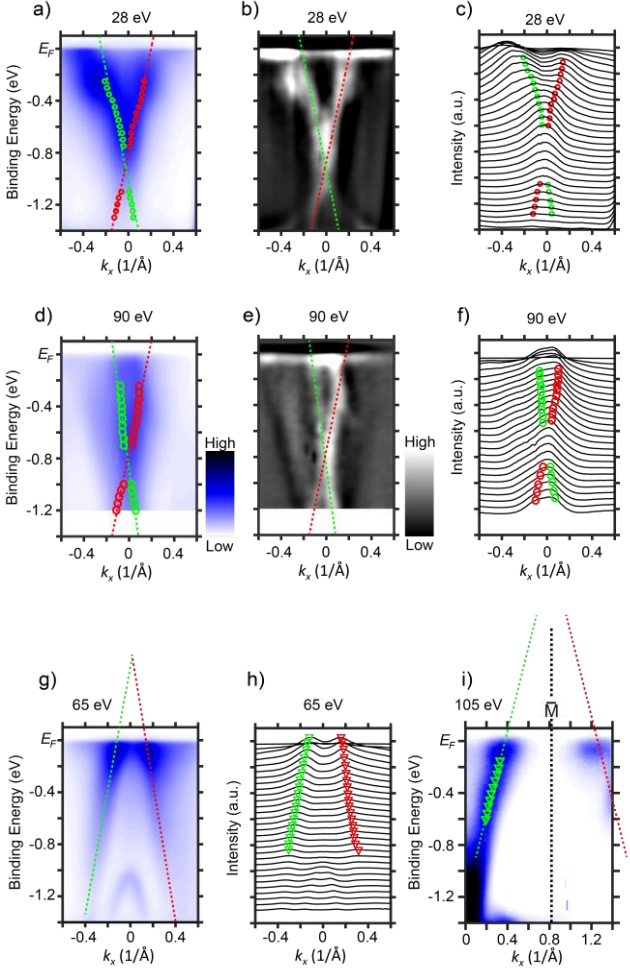

**Figure 4. MDC fitting of the linear dispersions at high symmetry points. a)-c)** ARPES spectra, second-derivative spectra and MDCs with photon energy of 28 eV, which corresponds to the bulk Γ point, respectively. **d)-f)** ARPES spectra, second-derivative spectra and MDCs with photon energy of 90 eV, which corresponds to the bulk Γ point in the next BZ in the $k_z$ direction, respectively. The green and red circles are the peaks extracted from MDCs. The red and green dotted lines are the linear fittings of the extracted MDC peak positions. **g) & h)** ARPES spectra and MDCs with photon energy of 65 eV at the $\bar{\Gamma}$ point of the surface BZ, respectively. The red and green triangles in h) are the peaks extracted from MDCs, and the red and green dotted lines in g) are the linear fittings of the MDC peak positions. **i)** ARPES spectra with photon energy of 105 eV near the $\bar{M}$ point of surface BZ. The black dotted line indicate the position of the $\bar{M}$ point of surface BZ. The green triangles and dotted lines are the peaks from the MDC fitting, respectively.

The topological band inversions of SnTe(111) are located at the L points of bulk BZ. Projected in [111] direction, this band inversions will induce topological surface Dirac cones at the $\bar{\Gamma}$ and $\bar{M}/\bar{M}'$ points in (111) surface. In our experiments, the Dirac cone at the $\bar{\Gamma}$ point is the most visible at the photon energy 65 eV, and the Dirac cones at the $\bar{M}$ point is the most visible at the photon energy 105eV. The photon energies, at which these two type of Dirac cones are most visible, coincide with the ARPES scan lines close to the L points in the bulk BZ as shown in Figure 2t. These Dirac cones at the $\bar{\Gamma}$ and $\bar{M}/\bar{M}'$ points show no $k_z$ dependent dispersion in the $k_x$-$k_z$ constant energy maps in Figure 3, which demonstrates the nature of surface states originated from the band inversions at the L points of bulk BZ. The Fermi velocities of these Dirac cones are in the same orders of magnitude to the ones in previous reports [16, 17], and the value for the Dirac cone at the $\bar{\Gamma}$ point is slightly higher than that at the $\bar{M}$ point. According to previous calculations on $Pb_{1-x}Sn_xTe(111)$, this difference can be attributed to the different orientations of the constant-energy ellipsoids around different L points. While the ellipsoid of the L point on the top of bulk BZ is projected perpendicularly to the $\bar{\Gamma}$ point of surface BZ, the other ellipsoids of the L points on the side of bulk BZ are tilted to the projection direction ($k_z$[111]) to the $\bar{M}/\bar{M}'$ points of surface BZ [31]. In addition to the difference of Fermi velocity, the Dirac point at the $\bar{M}$ point shows a higher energy position than that of the $\bar{\Gamma}$ point. We notice that the Fermi surface map in Figure 2s resembles the calculated one on the (√3×√3)R30°-1Sn terminated surface of SnTe(111) [30]. In this calculation, the energy positions of the Dirac cones at the $\bar{M}/\bar{M}'$ points is obviously higher than that at the $\bar{\Gamma}$ point for the (√3×√3)R30°-1Sn surface due to the charge-transfer-induced filling of the surface states. Such differences of both Fermi velocity and Dirac energy between the $\bar{\Gamma}$ point and the $\bar{M}/\bar{M}'$ points Dirac cones have also been reported in previous ARPES measurement [16].

In addition to the Dirac cones at the $\bar{\Gamma}$ point and the $\bar{M}/\bar{M}'$ points, we observed another Dirac-cone-like band structure only when the photon energy crosses the bulk Γ point of the BZ. The Fermi velocity of this new Dirac cone is of the same order of magnitude as the one for the surface Dirac cones found at the $\bar{\Gamma}$ points at the surface BZ. However, the position of the Dirac energy is much lower to bring the Dirac point below the Fermi energy. We note that these Dirac cones, only appearing at the photon energies corresponding to the bulk $\Gamma$ point, has not been predicted and observed in previous theoretical and experimental studies. No theoretical or experimental studies so far have reported a band inversion that could be responsible for a topological nature of such Dirac-cone-like electronic structure. We hope that further studies would reveal the nature of this newly observed Dirac-like linear band dispersion. If proven topological, such Dirac cone could shift the topological classification of SnTe(111) from a typical TCI. However, whether our findings and consequent theoretical/experimental studies would reinforce or overturn the previous findings that classify SnTe as a TCI is not yet clear.

## 5. Conclusions

We have successfully grown SnTe(111) on $BaF_2$(111) substrate via MBE method. The RHEED patterns and the experimental FS clearly indicate the [111] orientation of SnTe. A wide-range photon-energy-dependent ARPES study allows us to observe the band structure of SnTe(111) crossing the multiple high symmetry points of bulk BZ. In addition to the Dirac cones originating from band inversions at the L points, we observed yet another Dirac-cone-like linearly dispersing band at the $\Gamma$ point of the bulk BZ, which has the Fermi velocity of the same order of magnitude as that of surface Dirac cones but with much lower Dirac energy position. Our study calls for a more extensive theoretical and experimental investigation around the possible new Dirac state in SnTe.

## Acknowledgements

The work at Nanjing University is supported by the Fundamental Research Funds for the Central Universities No. 020414380037. ALS is supported by the U. S. DoE, Office of Basic Energy Science, under contract No. DE-AC02-05CH11231. The Stanford Institute for Materials and Energy Sciences and Stanford University is supported by the U. S. DoE, Office of Basic Energy Sciences, under contract No. DE-AC02-76SF00515. The work at Oxford University is supported from a DARPA MESO project (No. 187N66001-11-1-4105). Max Plank POSTECH Center for Complex Phase Materials is supported by the Ministry of Science, ICT & Future Planning of Korea under Project No. NRF-2011-0031558.